\documentclass{iau}
\usepackage{graphicx}

\usepackage{hyperref}
\hypersetup{
    bookmarks=false,                                  
    pdftitle={Stellar Evolution Models of Young Stars: Progress and Limitations},                    
    pdfauthor={Gregory A. Feiden},                   
    colorlinks=true,                                 
    linkcolor=black,                                 
    citecolor=black,                                 
    urlcolor=blue                                    
}

\title[Young Stellar Evolution Models] 
{Stellar Evolution Models of Young Stars: \\ Progress and Limitations}

\author[Gregory A. Feiden]   
{Gregory A.~Feiden$^1$}

\affiliation{$^1$Department of Physics \& Astronomy, Uppsala University, SE-751 20 Uppsala, Sweden 
                 \\email: {\tt gregory.feiden@physics.uu.se}}

\pubyear{2015}
\volume{314}  
\pagerange{119--126}
\jname{Young Stars \& Planets Near the Sun}
\editors{J. H. Kastner, B. Stelzer, \& S. A. Metchev, eds.}
\begin{document}

\maketitle

\begin{abstract}
Stellar evolution models are a cornerstone of young star astrophysics, which necessitates that they yield accurate and reliable predictions of stellar properties. Here, I review the current performance of stellar evolution models against young astrophysical benchmarks and highlight recent progress incorporating non-standard physics, such as magnetic field and starspots, to explain observed deficiencies. While addition of these physical processes leads to improved agreement between models and observations, there are several fundamental limitations in our understanding about how these physical processes operate. These limitations inhibit our ability to form a coherent picture of the essential physics needed to accurately compute young stellar models, but provide rich avenues for further exploration. 
\end{abstract}

\firstsection 

\section{Introduction}
Stellar evolution models are essential for understanding the time dependence of astrophysical phenomena. Theory provides absolute ages of stellar systems, which are particularly consequential for young populations, where a variety of processes are on-going that lead directly to observed properties of older stars and stellar systems. Ages provided by young stellar models inform our understanding about the lifetimes of protoplanetary disks, timescales for the formation of giant planets, timescales for stellar angular momentum evolution, and the time evolution of stellar magnetic activity. Furthermore, young stellar ages provide information regarding the formation and thermal evolution of stars and sub-stellar objects. Given their role in providing constraints on a number of astrophysical processes, it is critical that stellar models provide accurate characterizations of stellar populations, further necessitating that models yield accurate properties of individual stars.

This review attempts to consolidate our current understanding about the performance of young stellar evolution models and highlight current efforts to improve their accuracy. Development of young star models has a rich history; however, it will not be possible to provide a comprehensive historical review. Several shorter reviews exist for the interested reader (\cite[e.g., Stahler 1988]{Stahler1988}). Nevertheless, much of what we know today is the result of over half a century of hard work and dedication of theorists and observers, alike.

\section{Current Performance}
Performance of stellar evolution models can be assessed using two complementary astrophysical benchmarks: color-magnitude diagrams and touchstone stars. Each reveals different information about the successes and shortcomings of stellar evolution models.

\subsection{Color-Magnitude Diagrams}
Color-magnitude diagrams provide extensive diagnostic information about the validity of stellar evolution models across a range of effective temperatures and luminosities (see Bell, this volume). 
The current state of modeling color-magnitude diagram morphologies is conflicting, but can be roughly divided into three sub-categories encompassing clusters or stellar associations younger than 20 Myr, those between 20 and 100 Myr, and those that are older than 100 Myr. For the youngest populations, there is evidence of a significant luminosity spread at constant color in color-magnitude diagrams, a spread that cannot be reproduced by a single standard stellar model isochrone (\cite[e.g., Hillenbrand 1997; Da Rio \etal\ 2010]{Hillenbrand1997, DaRio2010a}). Observational errors do not appear to be the source of the spread. Instead, spreads are indicative of either genuine age spreads of several Myr resulting from extended star formation processes or they are highlighting effects that result from physics not presently included in standard stellar models. The precise origin of luminosity spreads is unresolved (\cite[e.g., Jeffries 2012]{Jeffries2012}), but multiple mechanisms have been identified that are able to generate intrinsic spreads in color-magnitude diagrams without the need to invoke extended star formation events. These include mechanisms like starspots (Somers, this volume) and episodic accretion (\cite[Baraffe \etal\ 2010]{Baraffe2010}).

Beyond about 20 Myr, the luminosity scatter in observed color-magnitude diagrams starts to decrease with  single standard stellar model isochrones providing more accurate representations of the data. This may simply be due to the fact that relative age differences of several Myr grow increasingly irrelevant after about 20 Myr or it may be the result of non-standard physics playing less of a role in governing the observed properties of stars. For example, the effects of early episodic accretion are only expected to be noticeable up to ages of around 20 Myr before theoretical predictions converge with standard stellar model results. Alternatively, it may be a combination of the two effects. However, this is contrasted by the fact that, for ages below about 100 Myr, there are noticeable disagreements between ages of young stellar populations derived from cool low-mass stars and those from hotter intermediate-mass stars that have reached the main sequence, with ages decreasing as stellar mass decreases (\cite[e.g., Hillenbrand 1997; Mamajek \& Bell 2014; Herczeg \& Hillenbrand 2015]{Hillenbrand1997, Mamajek2014, Herczeg2015}). So, while apparent age scatters at a given color decrease, standard models are unable to provide consistent age estimates across the color-magnitude diagram for young stellar populations, strongly suggesting that there are missing physics in stellar models. 

\begin{figure}[ht]
    \centering
    \includegraphics[width=0.47\linewidth]{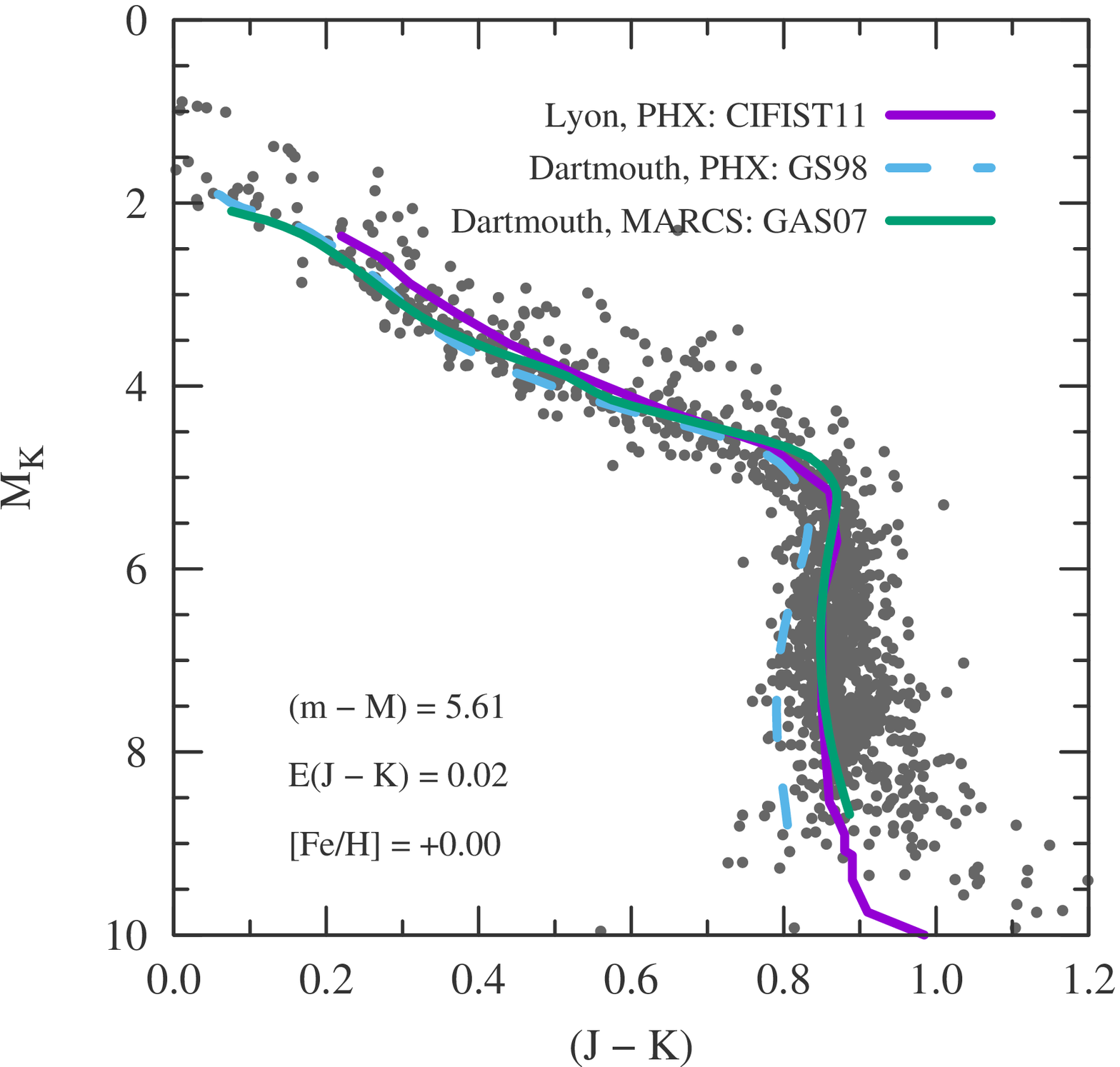} \qquad
    \includegraphics[width=0.47\linewidth]{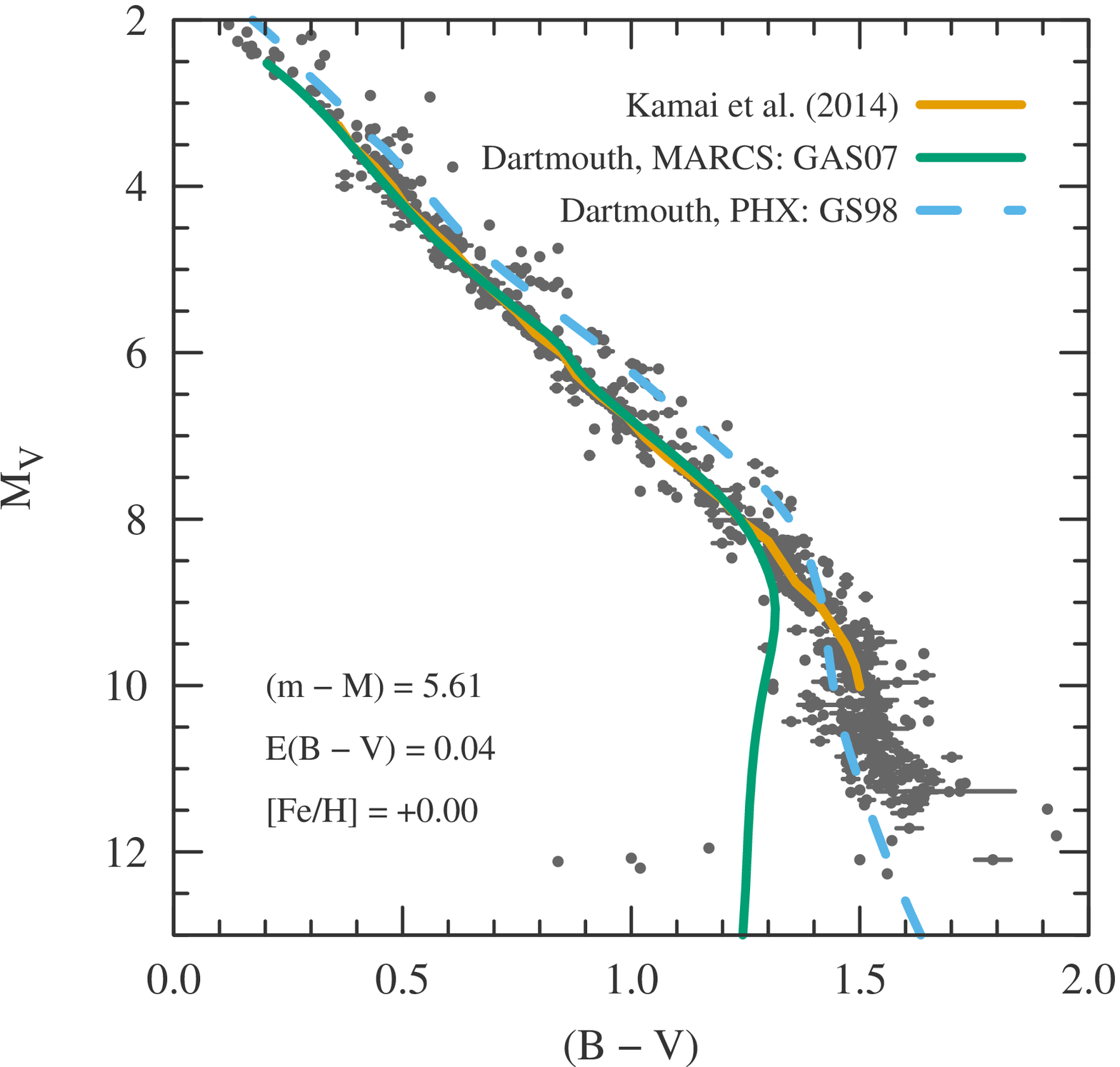}
    \caption{Color-magnitude diagrams for the Pleiades. Data are from \cite{Stauffer2007} and \cite{Kamai2014}. ({\it left}) 2MASS $(J-K)$-$M_K$. ({\it right}) Johnson $(B-V)$-$M_V$. Dartmouth (Feiden \etal, in prep) and Lyon (\cite[Baraffe \etal\ 2015]{Baraffe2015}) stellar evolution isochrones are overplotted. Dartmouth isochrones are shown with different solar abundances and bolometric corrections. An empirical isochrone from \cite{Kamai2014} is also shown in the right panel.}
    \label{fig:pleiades}
\end{figure}

For ages older than 100 Myr, a majority of the stars in young stellar populations are either on the main-sequence or their position in color-magnitude diagrams is indistinguishable from the zero-age main-sequence. The Pleiades provides a characteristic cluster sequence upon which to assess model validity for ages of approximately 100 Myr. The Pleiades has long been a source of trouble for stellar models of low-mass stars, with cooler K-dwarfs appearing to be significantly bluer than model and empirical predictions (\cite[Herbig 1962; Stauffer \etal\ 2003]{Herbig1962,Stauffer2003}). However, current generations of stellar models provide reasonable agreement across various color-magnitude diagrams, as demonstrated in Figure~\ref{fig:pleiades}.
Models yield excellent agreement in near-infrared (NIR) and optical color-magnitude diagrams until the onset of strong TiO absorption, signalling the beginning of the M dwarf sequence around $M_V = 8.0$ mag. Notably, a single standard model isochrone provides a consistent fit to the color-magnitude diagrams, in contrast with results from younger populations.

Furthermore, agreement is consistent across different flavors of models, as seen in Figure~\ref{fig:pleiades}. This improvement appears to be largely related to the adoption of updated model atmospheres used to prescribe bolometric corrections, which adopt recent solar abundances (\cite[Baraffe, this volume; Grevesse \etal\ 2007; Caffau \etal\ 2011]{Grevesse2007, Caffau2011}). In particular, lower oxygen abundances decrease the amount of water absorption in the NIR, yielding better agreement with NIR colors and providing support for their adoption in the face of known problems with helioseismology (\cite[Basu \& Antia 2004]{Basu2004}). Line lists and oscillator strengths for TiO and various monohydrides appear to still be a problem for the coolest models in optical passbands. As may be expected, optical colors are still fairly unreliable when it comes to drawing conclusions about young M stars.

Unfortunately, despite the utility of color-magnitude diagrams, they can mask significant errors present in stellar models. Bolometric corrections and photometric colors derived from either theoretical model atmosphere calculations or empirical methods can provide a counter-balance against more fundamental model errors.

\subsection{Stellar Fundamental Properties} 
Observations of touchstone (or benchmark) stars allow tests of stellar evolution theory at the most fundamental level by providing direct measurements of stellar masses, radii, effective temperatures, and luminosities. Dynamical systems with measured stellar masses are of greatest interest, as mass is a direct input parameter in stellar models, largely determining the entire evolution of a star. Over the past decade, several studies have leveraged dynamical information from young binary systems to confront stellar model predictions (\cite[e.g., Hillenbrand \& White 2004; Mathieu \etal\ 2007; Stassun \etal\ 2014]{Hillenbrand2004, Mathieu2007, Stassun2014}). Each study has invariably led to the same result: modern stellar evolution models are unable to accurately predict the properties of young, low-mass stars in binary systems. 

\cite{Hillenbrand2004} demonstrated that young stellar models systematically predict masses that are too small for a given stellar effective temperature and luminosity (radius). Their study relied on a large collection of predominantly astrometric binaries, with a few known eclipsing binary systems. Recently, \cite{Stassun2014} showed that the latest generations of stellar models fair no better than those adopted by \cite{Hillenbrand2004}. Using a smaller sample of carefully selected eclipsing binaries, stellar models were shown to overestimate observed masses for primaries, but underestimate masses for secondary stars. Models performed better above 1~$M_{\odot}$ than they did below that threshold. Still, models were unable to provide a mean absolute accuracy better than 5\% at high masses and no better than 20\% accuracy at low masses. 

However, one should be concerned about using binary systems to characterize the reliability of young stellar models. Young binary systems are thought to endure numerous strong dynamical interactions during their first few million years, potentially slowing stellar contraction rates by pumping energy into stars. In fact, there is a tantalizing correlation between the presence of significant modeling errors and the presence of a tertiary companion among eclipsing binary systems (\cite[Stassun \etal\ 2014]{Stassun2014}), suggesting  that tidal heating may be important when considering the structure and evolution of young stars.

Nevertheless, investigations focusing on samples of single stars with well determined temperatures and luminosities find internal inconsistencies in stellar models. As with color-magnitude diagrams, comparisons of stellar populations in the theoretical plane show that models predict systematically younger ages with decreasing effective temperatures (\cite[e.g., Malo \etal\ 2014]{Malo2014}). This may reflect discrepancies in color-magnitude diagrams being translated to the theoretical plane with the adoption of empirical bolometric corrections. However, \cite{Malo2014} relied on an independent method to arrive at the same conclusion, suggesting the result is a general feature of stellar model predictions.

\section{Progress \& Fundamental Limitations}
Efforts are currently underway to mitigate the aforementioned discrepancies between young stellar models and observations. These efforts have largely, though not exclusively, focused on magnetic fields and starspots.

\subsection{Magnetic Fields}
Stellar evolution models that include effects related to the presence of globally pervasive magnetic fields (\cite[D'Antona \etal\ 2000; Mullan \& MacDonald 2001; Feiden \& Chaboyer 2012]{DAntona2000,MM01,FC12b}) were recently applied to young stellar systems (\cite[MacDonald \& Mullan 2010; Malo \etal\ 2014]{MM10,Malo2014}). Magnetic fields can inhibit large-scale convective flows, decreasing a star's net outward flux of energy. For young stars that are powered by the release of gravitational potential energy, decreasing their convective energy flux cools their effective temperature and traps energy within their interiors, slowing their contraction. Magnetic young stars are therefore cooler and larger at a given age than their non-magnetic counterparts, but note that radii of young stars are not ``inflated'' in the same sense as radii of low-mass main sequence stars; magnetic young stars simply undergo a more gradual contraction.

Magnetic inhibition of convection provides a natural explanation for observed age differences as a function of effective temperature in young stellar associations (\cite[see, e.g., MacDonald \& Mullan 2010; Malo \etal\ 2014]{MM10,Malo2014}). At the same time, decreasing the effective temperature of young stars leads to an overall decrease in internal temperatures owing to their largely adiabatic stratification. Lithium depletion timescales are extended as a result, since it takes longer for stellar interiors to reach temperatures required for lithium destruction (\cite[$\sim 2.5$ MK; Malo \etal\ 2014]{Malo2014}), leading to better agreement between ages derived from the location of the lithium depletion boundary and other methods. Notably, \cite[MacDonald \& Mullan (2010; and later Malo \etal\ 2014)]{MM10} showed that the apparent age of 5 Myr derived for the low-mass stars in the $\beta$-Pic moving group can be reconciled with the 20 -- 30 Myr age estimate from the higher mass population, lending some support to the idea that magnetic fields are important in early phases of stellar evolution.

It is encouraging to note that magnetic models of young stars can reconcile ages of low-mass stars with the higher mass population using seemingly reasonable magnetic field properties. Surface magnetic field strengths are $\sim 3$ kG (equipartition values in low-mass K and M stars), while interior field strengths are $\sim 50$ kG. As tantalizing as magnetic fields may be, their adoption must be approached with caution. Significant uncertainties exist about what constitutes realistic interior and surface magnetic field strengths for young stars, particularly as a function of age and effective temperature. Nevertheless, magnetic models provide firm predictions about surface magnetic field strengths that require validation.

\subsection{Starspots}
An alternative to global magnetic inhibition of convection is that starspots block outgoing flux, causing stars to thermally restructure (\cite[Spruit \& Weiss 1986]{Spruit1986}; \cite[Jackson \etal\ 2009]{Jackson2009}). While starspots are, in part, the physical manifestation of inhibited convection on the surface of stars, their effects on stellar positions in color-magnitude diagrams are distinct. Global inhibition of convection leads to an overall shift of theoretical predictions toward redder colors, primarily due to their cooler effective temperatures. In contrast, starspots create a scenario where there are regions of different temperatures on the stellar surface, leading to emergent spectra and colors that are a combination of flux from various surfaces. Starspots can shift stars toward either bluer or redder colors as a result, depending on the precise areal coverages, temperature contrasts, and photometric colors under consideration (\cite[Jackson \& Jeffries 2014]{Jackson2014a}).

\cite{Jackson2009} demonstrate that spotted models can match observed morphologies of color-magnitude diagrams and fundamental properties of young low-mass stars. In particular, redistribution of flux throughout the stellar convection zone leads to unspotted regions of a star appearing hotter than for the same star without spots. This provides a natural explanation for the observed blue color of Pleiades K dwarfs and excess flux in blue regions of K dwarf spectra (\cite[Stauffer \etal\ 2003]{Stauffer2003}). Since spots also block outgoing flux, they can slow the contraction rate of young stars in much the same way as magnetic inhibition of convection (\cite[Jackson \&  Jeffries 2014]{Jackson2014a}). Finally, starspots may provide an explanation of observed spreads in color-magnitude diagrams and spreads in measured lithium abundances in young stellar populations (see Somers, this volume).

However, as with magnetic fields, spotted models must be approached with caution. There is some empirical data yielding estimates of starspot areal coverages and temperature contrasts, but only for a few isolated cases, making the process of validating spotted model predictions difficult. Synthesizing data from several color-magnitude diagrams and photometric lightcurves is currently the best means of constraining these models. Still, starspot physics is not very well understood. It is not clear that flux trapped by any given starspot is redistributed throughout the convection zone before the star is able to radiate away that excess energy (\cite[e.g., Spruit 1982]{Spruit1982}). Nor is it understood how this redistribution occurs: is flux completely redistributed without a need for the star to restructure, or must a star thermally restructure? These are interesting questions that we are now beginning to address, at least in the context of stellar structure and evolution. Often ignored for simplicity, starspots may come to play an important role in our understanding of young, low-mass stars.

\begin{discussion}

\discuss{Chabrier}{My concern with invoking magnetic fields to explain shortcomings of fully convective stellar models is that interior magnetic field strengths need to be of order 1 MG or greater. Magnetic fields of such a magnitude will be buoyantly unstable and cannot be maintained by dynamo action. What are your thoughts about this?}

\discuss{Feiden}{I agree that 1 MG or stronger magnetic fields deep within fully convective stars are probably not realistic and should be taken as a serious flaw in the models. However, this requirement is only necessary for fully convective \emph{main sequence} stars. Typical interior magnetic field strengths invoked to reconcile models with observations of low-mass pre-main-sequence stars are on the order of 50 kG, quite consistent with estimates of dynamo generated field strengths from 3D MHD simulations.}

\discuss{Pinsonneault}{Two points. First, adoption of the revised solar abundances not only introduces disagreement with the solar convection zone boundary, but predictions of neutrino rates are lower due to the low Fe abundance, although still formally consistent with observations. Second, we showed that not only did the Pleiades K dwarfs appear bluer in the CMD, but their spectra showed noticeable excess emission at bluer wavelengths. How would you explain this excess emission if theoretical colors are correct?}

\discuss{Feiden}{Interesting question. Some of this may be the result of the metallicity difference between stars in Praesepe and the Pleiades. Praesepe stars may exhibit more significant line blanketing in the blue, leading to the appearance of sizable continuum excess at blue wavelengths for Pleiades stars. Other emission, say from Balmer lines, may just not contribute significantly to the integrated flux, thus playing a negligible role in governing broadband optical colors. This is testable with synthetic spectra.}

\end{discussion}


\begin{thebibliography}{}

\bibitem[{{Baraffe} \& {Chabrier} (2010)}]{Baraffe2010}
{Baraffe}, I. \& {Chabrier}, G. 2010, \textit{A\&A}, 521, A44

\bibitem[{{Baraffe} \etal\ (2015)}]{Baraffe2015}
{Baraffe}, I., {Homeier}, D., {Allard}, F., \& {Chabrier}, G. 2015, \textit{A\&A}, 577, A42

\bibitem[{{Basu} \& {Antia} (2004)}]{Basu2004}
{Basu}, S. \& {Antia}, H.~M. 2004, \textit{ApJL}, 606, L85

\bibitem[{{Caffau} \etal\ (2011)}]{Caffau2011}
{Caffau}, E., {Ludwig}, H.-G., {Steffen}, M., {Freytag}, B., \& {Bonifacio}, P. 2011, \textit{SoPh}, 268, 255

\bibitem[{{Da Rio} \etal\ (2010)}]{DaRio2010a}
{Da Rio}, N., {Robberto}, M., {Soderblom}, D.~R., \etal\ 2010, \textit{ApJ}, 722,
  1092

\bibitem[{{D'Antona} \etal\ (2000)}]{DAntona2000}
{D'Antona}, F., {Ventura}, P., \& {Mazzitelli}, I. 2000, \textit{ApJ}, 543, L77

\bibitem[{{Feiden} \& {Chaboyer} (2012)}]{FC12b}
{Feiden}, G.~A. \& {Chaboyer}, B. 2012, \textit{ApJ}, 761, 30

\bibitem[{{Grevesse} \etal\ (2007)}]{Grevesse2007}
{Grevesse}, N., {Asplund}, M., \& {Sauval}, A.~J. 2007, \textit{SSRv}, 130, 105

\bibitem[{{Herbig} (1962)}]{Herbig1962}
{Herbig}, G.~H. 1962, \textit{ApJ}, 135, 736

\bibitem[{{Herczeg} \& {Hillenbrand} (2015)}]{Herczeg2015}
{Herczeg}, G.~J. \& {Hillenbrand}, L.~A. 2015, \textit{arXiv}: 1505.06518

\bibitem[{{Hillenbrand} (1997)}]{Hillenbrand1997}
{Hillenbrand}, L.~A. 1997, \textit{AJ}, 113, 1733

\bibitem[{{Hillenbrand} \& {White} (2004)}]{Hillenbrand2004}
{Hillenbrand}, L.~A. \& {White}, R.~J. 2004, \textit{ApJ}, 604, 741

\bibitem[{{Jackson} \& {Jeffries} (2014)}]{Jackson2014a}
{Jackson}, R.~J. \& {Jeffries}, R.~D. 2014, \textit{MNRAS}, 441, 2111

\bibitem[{{Jackson} \etal\ (2009)}]{Jackson2009}
{Jackson}, R.~J., {Jeffries}, R.~D., \& {Maxted}, P.~F.~L. 2009, \textit{MNRAS}, 339, L89

\bibitem[{{Jeffries} (2012)}]{Jeffries2012}
{Jeffries}, R.~D. 2012, in Star Clusters in the Era of Large Surveys, ed.
  A.~{Moitinho} \& J.~{Alves}, 163

\bibitem[{{Kamai} \etal\ (2014)}]{Kamai2014}
{Kamai}, B.~L., {Vrba}, F.~J., {Stauffer}, J.~R., \& {Stassun}, K.~G. 2014, \textit{AJ}, 148, 30

\bibitem[{MacDonald \& Mullan (2010)}]{MM10}
MacDonald, J. \& Mullan, D.~J. 2010, \textit{ApJ}, 723, 1599

\bibitem[{{Malo} \etal\ (2014)}]{Malo2014}
{Malo}, L., {Doyon}, R., {Feiden}, G.~A., \etal\ 2014, \textit{ApJ}, 792, 37

\bibitem[{{Mamajek} \& {Bell} (2014)}]{Mamajek2014}
{Mamajek}, E.~E. \& {Bell}, C.~P.~M. 2014, \textit{MNRAS}, 445, 2169

\bibitem[{Mathieu \etal\ (2007)}]{Mathieu2007}
Mathieu, R.~D., Baraffe, I., Simon, M., Stassun, K.~G., \& White, R. 2007, in
  Protostars \& Planets V, 411

\bibitem[{Mullan \& MacDonald (2001)}]{MM01}
Mullan, D.~J. \& MacDonald, J. 2001, \textit{ApJ}, 559, 353

\bibitem[{{Spruit} (1982)}]{Spruit1982}
{Spruit}, H.~C. 1982, \textit{A\&A}, 108, 348

\bibitem[{{Spruit} \& {Weiss} (1986)}]{Spruit1986}
{Spruit}, H.~C. \& {Weiss}, A. 1986, \textit{A\&A}, 166, 167

\bibitem[{{Stahler} (1988)}]{Stahler1988}
{Stahler}, S.~W. 1988, \textit{ApJ}, 332, 804

\bibitem[{{Stassun} \etal\ (2014)}]{Stassun2014}
{Stassun}, K.~G., {Feiden}, G.~A., \& {Torres}, G. 2014, \textit{New Astronomy Reviews}, 60, 1

\bibitem[{{Stauffer} \etal\ (2007)}]{Stauffer2007}
{Stauffer}, J.~R., {Hartmann}, L.~W., {Fazio}, G.~G., \etal\ 2007, \textit{ApJS},
  172, 663

\bibitem[{{Stauffer} \etal\ (2003)}]{Stauffer2003}
{Stauffer}, J.~R., {Jones}, B.~F., {Backman}, D., \etal\ 2003, \textit{AJ}, 126, 833

\end{thebibliography}
\end{document}